\documentclass[aps,onecolumn,12pt,preprint]{revtex4-1}
\usepackage{graphicx}
\usepackage{amssymb,amsmath}
\usepackage[usenames]{color}
\newcommand{\pslash}{{p\hspace{-5pt}/}}

\newcommand{\epslash}{{\epsilon \hspace{-5pt}/}}
\newcommand{\del}{\partial}

\begin{document}
\title{A Hidden Local Symmetry Approach to Rho Meson Photoproduction}
\author{Hiromi \surname{Kaneko}}
\email{kanekoh@rcnp.osaka-u.ac.jp}
\author{Atsushi \surname{Hosaka}}
\email{hosaka@rcnp.osaka-u.ac.jp}
\affiliation{Research Center for Nuclear Physics, Osaka University, Mihogaoka 10-1, Osaka 567-0047, Japan}
\author{Olaf \surname{Scholten}}
\email{scholten@kvi.nl}
\affiliation{Kernfysisch Versneller Instituut, University of Groningen, 9747 AA, Groningen, The Netherlands}
\date{\today}
\begin{abstract}
We study photoproduction of $\rho$ mesons in a model of hidden local symmetry.
 The $\rho$ meson is introduced as a hidden gauge boson and  a 
 phenomenological $\rho$ meson-nucleon Lagrangian is constructed
 respecting chiral symmetry.
 It is shown that the $\sigma$ exchange
 interaction is needed for neutral $\rho$ meson
 photoproduction to reproduce the experimental cross sections. For
 charged $\rho$ meson photoproduction, the model takes into account
 the $\rho$ meson magnetic moments from the three-point vertex in the
 kinetic terms. We show that the magnetic moment of the charged $\rho$ meson has
 a significant effect on the total cross sections
through the $\rho$ meson exchange process,
which is proportional to the energy of the photon.
The $t$-channel dominance may be used for the study of structures of
 various unstable particles.
\end{abstract}
\pacs{}
\keywords{rho meson photoproduction, hidden local symmetry, magnetic moments}
\maketitle
\section{Introduction}
Vector mesons play important roles in hadron physics. For instance, they
are responsible for the short range part of the nuclear force and also
explain electromagnetic properties of hadrons through the vector meson
dominance model~\cite{Nambu:1997vw,Sakurai196906,Machleidt:1987hj}.
They have been studied in the quark model as $q \bar q$ states, while in
field theoretical approaches they are regarded as gauge bosons of
certain gauge symmetries.
The $\rho$ meson has been introduced as a gauge boson of
the Hidden Local Symmetry (HLS) of the non-liner $\sigma$ model which has
explained many celebrated low energy
relations~\cite{Bando:1984ej,Bando:1987br}.
Also, a holographic model in the string theory provides a systematic
derivation of the series of vector
mesons~\cite{Sakai:2004cn,Sakai:2005yt}.
Another recent development is to regard a vector meson as one of
the building blocks of the so called molecular-resonance states.
Some studies have been done for the $\rho$-N(nucleon) systems and their
extensions~\cite{Roca:2005nm,Khemchandani:2011et,Khemchandani:2011mf}.
They are considered as candidates of
non-conventional multi-quark states in the quark model. Such resonant
states may be considered as strongly correlated three-body states due to
the $\rho$ meson decay into two pions.
\par
Since vector mesons are unstable, only production reactions can provide
information on their structure and interaction with the other hadrons.
Among them, photoproduction is the most
useful because, through vector meson dominance, the real photon may
convert into a virtual vector meson which successively interacts with
the nucleon. 
Photoproduction of a neutral $\rho$ meson can easily be measured using the decay of the $\rho^0$
into two charged pions $\pi^+ \pi^-$~\cite{Wu:2005wf,:1968ke}.
\par
In this paper, we investigate $\rho$ meson photoproduction for the
study of the $\rho$ meson dynamics based on the symmetries of the strong
interaction.
For this purpose, we employ the HLS model.
It turns out that this model by itself can not explain the production rate as
observed in experiments. Therefore, we follow previous
work~\cite{Friman:1995qm} and introduce the $\sigma$ meson exchange in a
phenomenological manner. We also investigate charged $\rho$ meson
production where we propose a method to study the unknown
magnetic moment of the $\rho$ meson. This is possible if the reaction
mechanism is dominated by $\rho$ meson exchange in the
$t$-channel as expected in the forward region.
\section{Model}
Let us introduce the HLS Lagrangian based on the non-linear $\sigma$ model with
the vector mesons, which is given by
\begin{gather}
{\mathcal{L}}_{\text{HLS}} = - \frac{1}{2} \text{Tr} F_{\mu \nu}^2
+ a {\mathcal{L}}_{V} + {\mathcal{L}}_{A}\, ,
\label{L_HLS}
\end{gather}
where
\begin{align}
{\mathcal{L}}_{V} &=
- \frac{f_\pi^2}{4} \text{Tr} \left[
(\del_\mu \xi^\dagger - ig V_\mu \xi^\dagger + ie\xi^\dagger A_\mu) \xi
+ ( \xi \leftrightarrow \xi^\dagger) \right]^2 \, ,
\label{L_V}
\\
{\mathcal{L}}_{A} &=
 - \frac{f_\pi^2}{4} \text{Tr} \left[
(\del_\mu \xi^\dagger - ig V_\mu \xi^\dagger + ie\xi^\dagger A_\mu) \xi
-( \xi \leftrightarrow \xi^\dagger) \right]^2\, .
\label{L_A}
\end{align}
Here we follow the notations for the fields and the normalization from
Ref.~\cite{Bando:1984ej}. For instance, the isospin structure is given by
\begin{gather}
 V^{\mu}=\frac{\vec{\tau}}{2}\cdot \vec{V}^{\mu},\,
  A^{\mu}\to\frac{\tau^3}{2}A^{\mu},\,
  \pi=\frac{\vec{\tau}}{2}\cdot\vec{\pi}~\, \text{and}~\, \xi=\exp\left(i\pi/f_\pi\right),
\end{gather}
with $f_\pi$ being the pion decay constant.
We also include, in addition to the $\rho$ meson field $V^\mu$, the photon field $A^\mu$.
The anomalous term generates the isoscalar photon coupling to the $\omega$ meson.
Then the relevant terms for photoproduction are
\begin{align}
{\mathcal{L}}_{V\gamma} &= -e\frac{m_{V}^2}{g}A^{\mu}\left(\rho^3_{\mu}+\frac{1}{3}\omega_{\mu}\right)\, ,
\label{L_V-gamma}\\
 {\mathcal{L}}_{VVV} &= -2ig \text{Tr}\left(\partial_{\mu}\rho_{\nu}\left[\rho^{\mu},\rho^{\nu}\right]\right)\, ,
 \label{L_VVV}\\
 {\mathcal{L}}_{\omega\pi\rho}
 &= g_{\omega\rho\pi}\epsilon_{\mu\nu\rho\sigma}\partial^{\mu}\vec{\rho}^{~\nu}\partial^{\rho}\omega^{\sigma}\vec{\pi}\, .
\label{L_omega-pi-rho}
\end{align}
The $\omega$ meson coupling is given by
$g_{\omega\rho\pi}=-3g^2/8\pi^2f_{\pi}$~\cite{Fujiwara:1984mp}.
\par
\begin{figure}[htbp]
 \begin{center}
  \includegraphics[scale=0.7]{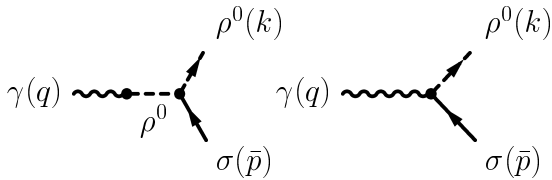}
 \end{center}
 \caption{The vertices of $\gamma\sigma\rho$ in the hidden local symmetry approach.}
 \label{fig.sigmarho}
\end{figure}

As far as neutral $\rho$ meson photoproduction is concerned, it turns
out that the above ingredients are not enough to reproduce the
experimental data using these interactions. To explain the observed
strength, the scalar isoscalar $\sigma$ meson was introduced in
Ref.~\cite{Friman:1995qm}. We adopt the phenomenological Lagrangian
given by
\begin{align}
 {\mathcal{L}}_{\sigma\rho\rho} &= \frac{g_{\sigma\rho\rho}}{m_{\rho}}\sigma\left(\partial_{\mu}\rho_{\nu}^{0}\partial^{\mu}\rho^{\nu 0}-\partial_{\mu}\rho_{\nu}^{0}\partial^{\nu}\rho^{\mu 0}\right)\, .
\label{L_sigma-rho-rho}
\end{align}
For photoproduction off a nucleon, $\rho NN$, $\sigma NN$ and $\pi NN$ interaction vertices are needed:
\begin{align}
  {\mathcal{L}}_{\rho NN} &= -g_{\rho NN}\bar{N}\left(\gamma_{\mu}\rho^{\mu}+\frac{\kappa_{\rho}}{4M}\sigma^{\mu\nu}F_{\mu\nu}\right)N\, ,
\label{L_rhoNN}\\
  {\mathcal{L}}_{\sigma NN}&=  g_{\sigma NN}\sigma\bar{N} N\, ,
\label{L_sigmaNN}\\
{\mathcal{L}}_{\pi NN} &= -2ig_{\pi NN}\bar{N}\gamma_5\pi N\, .
 \label{L_piNN}
\end{align}
One would expect that the $\sigma$ meson could be incorporated in the
model by introducing the scalar fluctuation, $f_\pi\to f_\pi+\sigma$.
As shown in Fig.\ref{fig.sigmarho}, this produces two amplitudes
involving the $\sigma$-exchange, which however cancel exactly. For the $\pi NN$
vertex we have adopted the pseudo-scalar coupling, which is equivalent to
the pseudo-vector coupling for on-shell nucleons as is the case when the
$\pi$ appears in the $t$-channel. The Lagrangians
Eqs.~(\ref{L_V-gamma})--(\ref{L_sigmaNN}) complete all the necessary
interactions in the present analysis. The needed parameters are given in
Table~\ref{cup_para}.
\begin{table}[htbp]
\caption{Coupling constants for various interactions.}
\label{cup_para}
 \begin{tabular}{cccccccc}\hline\hline
 $e^2/4\pi$ & $g(=g_{\rho NN})$ & $f_{\pi}$(MeV)  & $g_{\pi NN}$ & $g_{\sigma NN}$ & $g_{\sigma\rho\rho}$ & $\kappa_\rho$ \\
 \hline
   1/137  & 5.85 &  93   &  13.26       & 10              & 11             & 3.7   \\
   \hline\hline
 \end{tabular}
\end{table}
\begin{table}[htbp]
\caption{ Masses and cutoff parameters in the present calculation (MeV).}
 \begin{tabular}{cccccccc}\hline\hline
 $m_{\pi}$ & $m_{\sigma}$ & $m_{\rho}$ & $m_{\omega}$ & $M_{N}$ &
  $\Lambda$ & $\Lambda_\sigma$ &$\Lambda_{\sigma\rho\rho}$\\ \hline
   137     & 500          &    770     &  782         & 938 & 800 & 900
			  & 700      \\
\hline\hline
 \end{tabular}
\label{mass_para}
\end{table}

\begin{figure}[htbp]
\begin{center}
\includegraphics[scale=0.7]{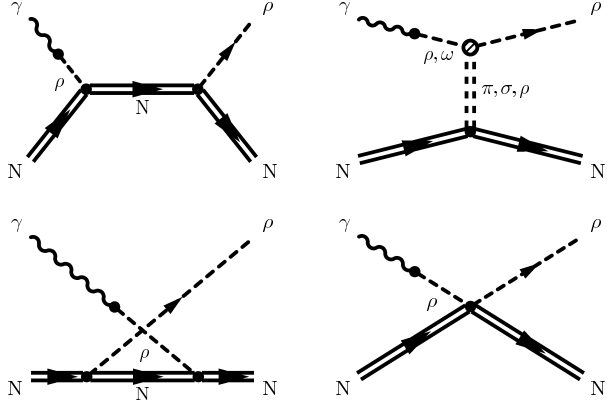}
 \caption{Feynman diagrams for the tree-level model: double line presents nucleon, wavy line is photon, dashed line is $\rho$ meson, double dashed line is $\pi$, $\sigma$ and $\rho$ mesons.}
\label{fig.tree-level}
\end{center}
\end{figure}

Having the above interactions, one can obtain the tree-level amplitude as a
sum of the $s$-channel, $t$-channel ($\pi$, $\sigma$, $\rho$ exchange),
$u$-channel and contact terms as shown in Fig.~\ref{fig.tree-level}.
The $s$-channel amplitude, for instance, is given by
\begin{align}
{\mathcal{M}}_s
&= -eg_{\rho\text{NN}}\bar{u}_{\text{p}}(p')\left(\epslash^*(k)+\frac{\kappa_{\rho}}{2M_N}\sigma_{\mu\nu}(ik^{\nu}\epsilon^{\mu*}(k))\right) \nonumber \\
&\times
\frac{i(\bar{\pslash}+M_N)}{\bar{p}^2-M_N^2}\left(\epslash^{\gamma}(q)+\frac{\kappa_{\text{p}}}{2M_N}\sigma_{\mu\nu}(-iq^{\nu}\epsilon^{\gamma}(q))\right)u_{\text{p}}(p)\, ,
\end{align}
where the vectors $\epsilon^\gamma_{\mu}$ and $\epsilon_{\mu}$ are
photon and $\rho$ meson polarization vectors.
Here the amplitude $\mathcal{M}_s$ is defined by the $S$-matrix through
$S = 1 +\mathcal{M}_s$.
In this calculation we introduce form factors following the
prescription of Davidson and Workman~\cite{Davidson:2001rk}. For all
($s$, $t$ and $u$) channels, the amplitudes are separated into
gauge-invariant and non gauge-invariant parts. The sum of the non
gauge-invariant parts is, however, gauge-invariant.
Therefore, we multiply the form factors $F_s, F_u$ and $F_t$ to the
corresponding gauge-invariant parts of $s$, $t$ and $u$-channels, and
$F_c$ (common form factor) to the sum of the non gauge-invariant parts.
We employ the $s,t,u$-channel form factors as
\begin{gather}
F_{x}=\frac{\Lambda^4}{\Lambda^4+(M^2_x-q)^2},\  x =s,t,u\, ,
\end{gather}
where $M_x$ is the mean of particle in the channel $x$,
while for the common form factor
\begin{gather}
F_c = F_s+F_t+F_u-F_sF_t-F_tF_u-F_sF_u+F_sF_tF_u\, .
\end{gather}
For the $\sigma$ exchange $t$-channel,
we employ a form factor as
\begin{gather}
F_{\sigma}=\frac{\Lambda_{\sigma}^2-m_{\sigma}^2}{\Lambda_{\sigma}^2-q^2}\frac{\Lambda_{\sigma\rho\rho}^2-m_{\sigma}^2}{\Lambda_{\sigma\rho\rho}^2-q^2}\, .
\end{gather}
The cutoff parameters in the form factors are given in
Table~\ref{mass_para}.
These values are consistent with a finite size of hadrons of about 0.5\,fm,
which is related to the cutoff parameter $\Lambda$  by
$r\sim\sqrt{6}/\Lambda$~\cite{ThomasWeise200105}.

\section{Results and discussion}
\begin{widetext}
\begin{figure}[htbp]
\makebox[\textwidth][c] {
\includegraphics[width=0.45\textwidth]{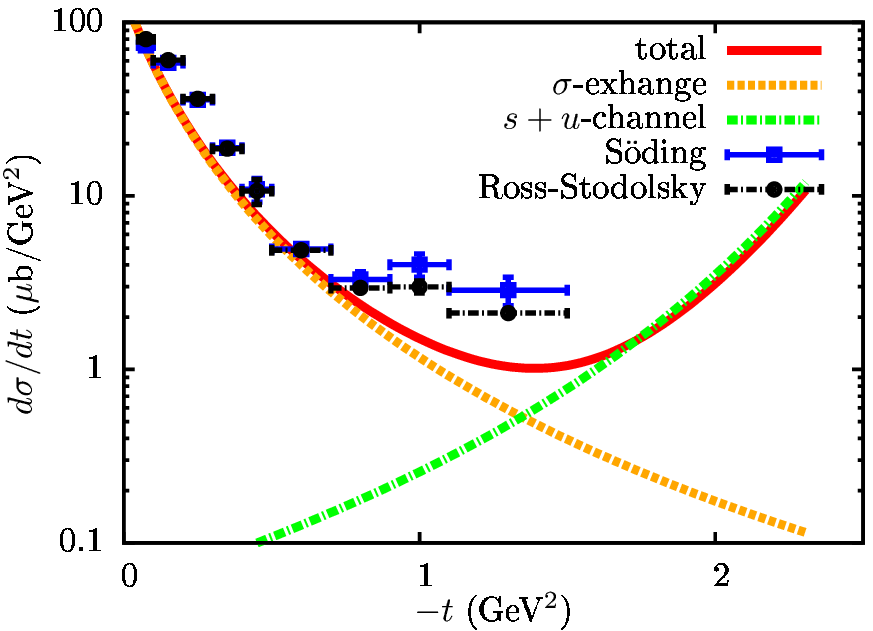}
 \hspace{0.05\textwidth}
\includegraphics[width=0.45\textwidth]{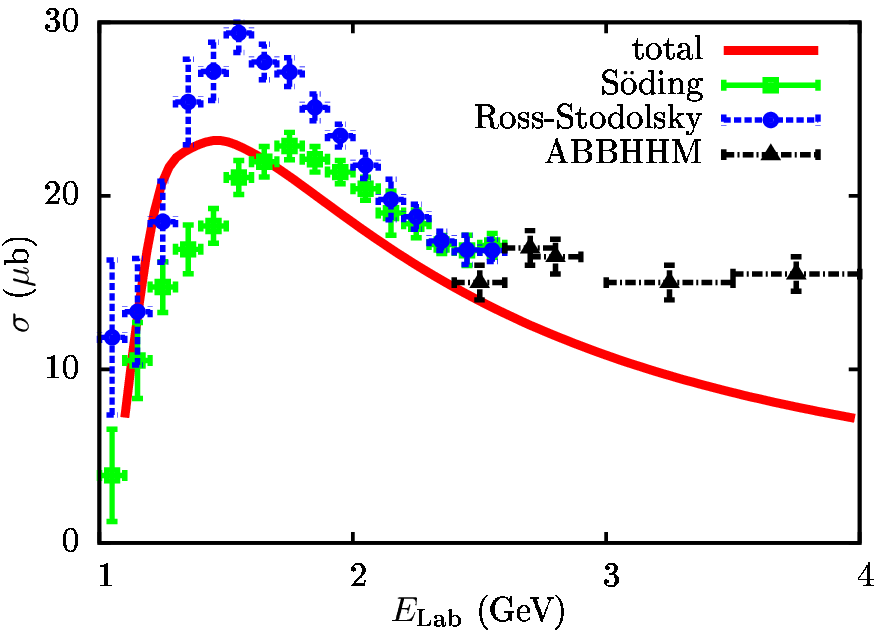}
}
\caption{Differential and total cross sections of the $\gamma p\to\rho^0 p$ reaction
as compared to the experimental data~\cite{Wu:2005wf,:1968ke,Soding:1965nh,Ross:1965qa}.
 The differential cross section is plotted as a function of the transferred momentum $t$
 at $E_{\gamma} = 2$~GeV.
}
\label{fig.rho0}
\end{figure}
\end{widetext}

First, we discuss differential and total cross sections for neutral
$\rho$ meson photoproduction in comparison with experimental data which
has been taken from Ref.~\cite{Wu:2005wf}. As anticipated in
Ref.~\cite{Friman:1995qm}, the phenomenological $\sigma$ meson exchange
in the $t$-channel plays a dominant role. It is consistent with the
observed data which shows strongly forward peaking as shown in
the left panel of Fig.~\ref{fig.rho0}. We also show various
contributions separately; we see that the $\sigma$ exchange dominates
except for the backward region, where the $u$-channel process becomes
important. The energy dependence of the total cross
section is shown in the right panel of Fig.~\ref{fig.rho0}. The present
calculation provides a smooth energy dependence; from the threshold it increases as the
phase space volume increases and then turns to decrease gradually above
$E_\gamma \sim 1.5$~GeV, partly due to the form factors.
On the contrary, experimental data shows a peak structure at
around $E_\gamma\sim1.5$ - $1.7$~GeV, depending on the
method of data analysis. This is perhaps due to nucleon
resonances which couples to the $\rho$ meson. Neglecting possible
resonance contributions, we can say that the agreement of the present
model with the experimental data is fair up to $E_\gamma \sim 3$~GeV.
Above this energy, the experimental data seems rather flat up to $4$~GeV, while the
present result keeps decreasing. The energy $E_\gamma \sim 3$~GeV is
already about $2$~GeV above the threshold, which is beyond the
limitation of the present model.
\par
Let us now turn to the charged $\rho$ meson photoproduction. For the
reaction $\gamma n \to \rho^- p$
the $\sigma$ meson exchange process 
is not allowed but instead charged $\rho$ meson exchange is.
In the HLS model, the $\gamma\to \rho^+ \rho^-$ amplitude is
given by the vector meson dominance $\gamma \to \rho^0$ and the
successive three-point vertex of $\rho^0 \rho^+ \rho^-$.
We find that in the $\rho^-$  photoproduction the $\rho$ meson exchange
process has the largest contribution, and therefore this process is
sensitive to the 
$\gamma \to \rho^+\rho^-$ vertex. 
Since the $\rho$ meson is a spin one particle there are three multipole components possible:
electric, magnetic and quadrupole couplings~\cite{Clark:1970xr}.
The strength of the electric coupling is unambiguously determined 
by the charge due to the gauge symmetry, while those of the
higher multipoles are not subject to the symmetry and can take an
arbitrary strength. 
At low energies the electric coupling, the lowest multipole, 
dominates but as the photon energy is increased the higher multipoles become more
important as they contain the photon momentum in their couplings. Here, for
simplicity, we 
consider only the effects of the lowest two multipoles. 
Ignoring the effects of the quadrupole moment corresponds to assuming
that spatial deformation of the $\rho$ meson is not large~\cite{Glozman:2011gf}.
\begin{figure}[htbp]
\begin{center}
\includegraphics[scale=0.8]{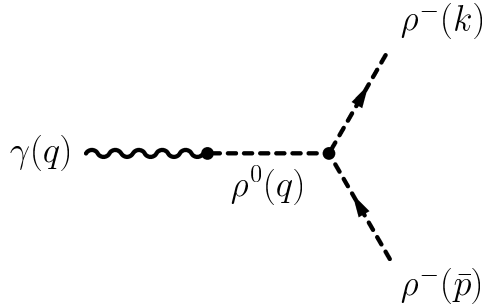}
\caption{The $\gamma\rho\rho$ vertices based on HLS model}
\label{fig.grr}
\end{center}
\end{figure}
\par
To illustrate the role of the magnetic moment of the $\rho$ meson,
let us look at the $\gamma \rho \rho$ vertex and its contribution to the
photoproduction through the $t$-channel $\rho$ meson exchange diagram(Fig.~\ref{fig.grr}).
The vertex is given as
\begin{align}
&\Gamma_{\text{HLS}}=ie\{2\epsilon^{\gamma}(q)\cdot k \epsilon^*(k)\cdot\epsilon(\bar{p})\notag \\
&+q\cdot\epsilon^{*}(k)\epsilon^{\gamma}(q)\cdot\epsilon(\bar{p})-k\cdot\epsilon(\bar{p})\epsilon^{\gamma}(q)\cdot\epsilon^*(k)\notag \\
&+q\cdot\epsilon^*(k)\epsilon^{\gamma}(q)\cdot\epsilon(\bar{p})-q\cdot\epsilon(\bar{p})\epsilon^{\gamma}(q)\cdot\epsilon^*(k)\}\, .
\label{eq.diff_mag}
\end{align}
The momentum $k,q$ and $\bar{p}$ are defined in Fig.~\ref{fig.grr}.
The first term is the electric coupling, while the third term is the
magnetic one.
The second term is similar to the magnetic coupling, and in fact, it
coincides the third term when the two $\rho$ mesons are on mass-shell.
Furthermore, one can verify that the contributions of the second and the third
terms to the photoproduction process
in the $t$-channel $\rho$ meson exchange diagram also coincide.
Therefore, the $t$-channel $\rho$ meson exchange contributions
are divided into the electric one with the first term of Eq.~(\ref{eq.diff_mag}),
and magnetic one with the second and the third terms.
Now the magnetic moment of the $\rho$ meson in the HLS model,
the sum of the two
terms of Eq.~(\ref{eq.diff_mag}), is given in the non-relativistic limit as
\begin{gather}
-\frac{\mu}{2m_V} \vec{q} \times \vec{\epsilon}^{\gamma} \cdot
 \vec{S}_\rho,\
\label{eq.mangetic_coupling}
\end{gather}
where $\mu = 2$ and $\vec S_\rho$ is the spin operator for the $\rho$ meson.
The factor $1/2m_V$ appears when the normalization of the $\rho$ meson wave function
is taken into account properly.
In general, the magnetic moment $\mu$ reflects the information of the internal
structure of the $\rho$ meson and can take any value.
Therefore, to see the role of the magnetic moment in photoproduction,
we treat $\mu$ as a parameter.
In the minimal gauge coupling of the photon to the $\rho$ meson
one has $\mu = 1$.
In the naive constituent quark model where the quark mass is half of the mass of the $\rho$, $m_q \sim m_V /2$,
$\mu = 2$, which coincides the value of the HLS model.
\par
\begin{figure}[htbp]
\begin{center}
\begin{widetext}
\makebox[\textwidth][c]{
\includegraphics[width=0.45\textwidth]{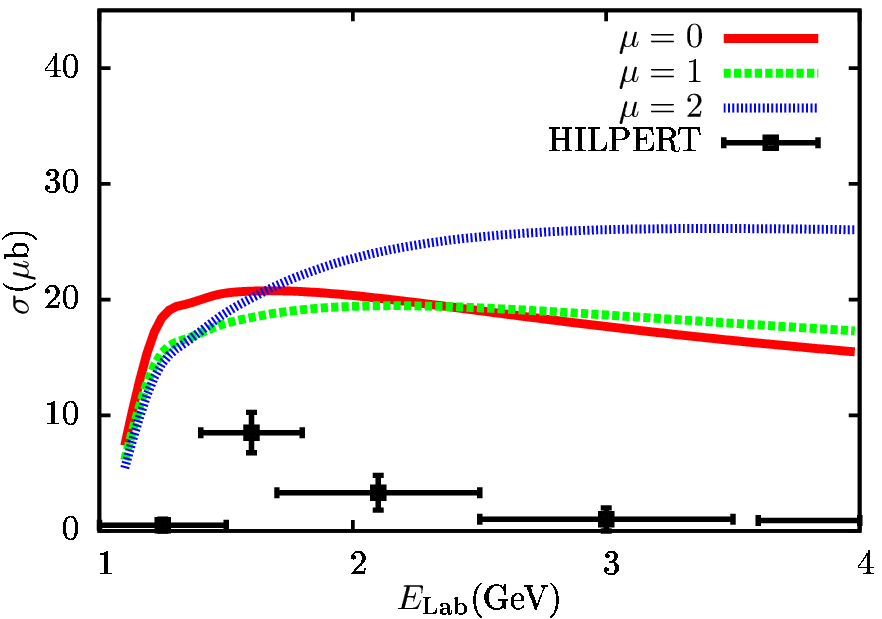}
\hspace{0.05\textwidth}
\includegraphics[width=0.45\textwidth]{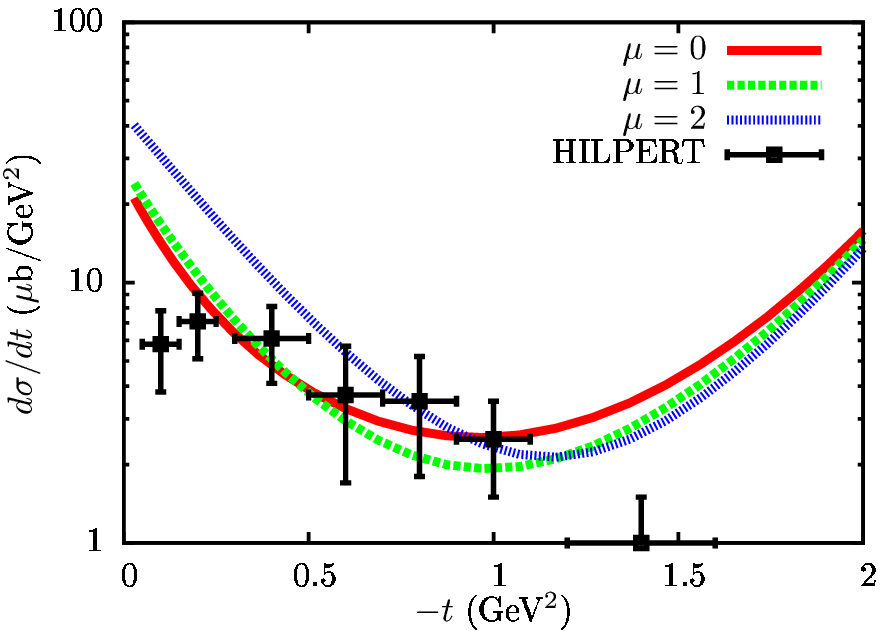}
}
\caption{Total and differential cross sections of the $\gamma n\to
 \rho^- p$ reaction with different value of $\mu=0,1,2$.
The experimental data are taken from Ref. ~\cite{Hilpert:1970jd} and the
 differential cross section is plotted as the transferred momentum $t$
 dependence at $E_{\gamma} = 2$~GeV.
The results for $\mu=2$ corresponds to those of the HLS model.}
\label{fig.rho-}
\end{widetext}
\end{center}
\end{figure}
In Fig.~\ref{fig.rho-}, we show the results for the total and
differential cross sections for magnetic moment values $\mu=0,1,2$
in units of $\rho$ meson magneton $e/2m_V$.
For the total cross section the agreement of our present study and
experimental data is not very good.
Indeed, the difference is increased ($E_\gamma\geqq 2 \text{GeV}$)
as the photon energy is increased as expected from the structure of
the magnetic coupling where
the photon momentum is involved.
For $\mu = 0$, the cross section is smallest,
since the only electric coupling terms exist.
With this value, the present calculation overestimates
the experimental data by about factor two near the threshold region,
$E_\gamma \sim 1.5$~GeV, while the data decrease
rapidly beyond that energy and the difference with the calculation increases.
If $t$-channel exchange mechanism dominates, however, this result
seems natural, because the cross section is roughly proportional
to $s^J$ where $J$ is the spin of the exchanged particle.
\par
Turning to the $t$-dependence (angular distribution), the $\rho$-exchange
process exhibits a strong forward peak as shown in the right panel of Fig.~\ref{fig.rho-}
irrespective of
the value of the magnetic moment $\mu$.
This again seems contradictory to experimental data, which
is rather flat for small $|t|$ and decreases for $|t|\geq 0.2$~GeV$^2$.
In the present calculation there are not many parameters
to adjust the energy and $t$ dependence of the cross section,
and therefore, our result for the $\rho$-exchange amplitude is
rather constrained.
\par
\begin{figure}[htbp]
\begin{center}
\begin{widetext}
\makebox[\textwidth][c]{
\includegraphics[width=0.5\textwidth]{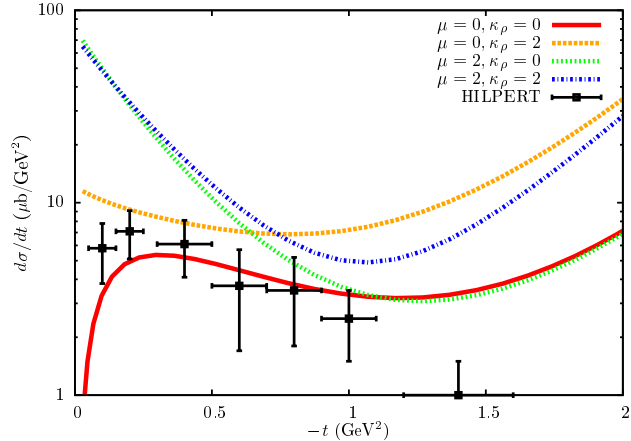}
}
\caption{Differential cross sections of the $\gamma n\to
 \rho^- p$ reaction with different value of $\mu=0,2$ and $\kappa_\rho=0,2$.
the differential cross section is plotted as the transferred momentum
 $t$ dependence at $E_{\gamma} = 2$~GeV and the cutoff parameter at $\Lambda = 1.1$~GeV.
}
\label{fig.rho-_test}
\end{widetext}
\end{center}
\end{figure}

To investigate the origin of the strong forward peaking predicted by our
calculation, we have checked the various contributions to the cross section.
This shows that the forward contribution is partly due to the magnetic
contribution of the $t$-channel $\rho$ meson exchange.
Another important contribution is from the magnetic (tensor) coupling of the
$\rho$ meson with the nucleon with the strength $\kappa_\rho$.
To investigate these points quantitatively we show in Fig.~\ref{fig.rho-_test}.
the results of calculations where we set the $\rho$ magnetic moment, $\mu$, equal to zero,
combined with setting $\kappa_\rho$ equal to zero.
As can be seen the vanishing cross section at forward angles is
reproduced only when both $\mu=0$ and $\kappa_\rho=0$, though there 
remains still a discrepancy at backward angles, where in
our calculation the $u$-channel process contributes significantly.
The null values of $\mu$ and $\kappa_rho$ are in clear contradiction with the predictions of the meson
chiral model.

\par
Our calculations form a strong motivation to improve the existing data
base for photoinduced charged $\rho$ meson production as this is very
sensitive to the magnetic moment.
Since a neutron target is not available for this work one has to perform
experiments on the deuteron for which the application of the minimum
spectator method is applied to obtain better kinematical constraints~\cite{Nakano:2008ee}.
In addition it is important to have more accurate data for $t \geq1.1~ \text{GeV}^2$.

\section{Conclusions}
We have studied photoproduction of a $\rho$ mesons based
on a chiral model with vector mesons introduced in the HLS model
supplemented by a $\sigma$ meson exchange.
The production of the neutral $\rho$ meson is reasonably explained by
the $\sigma$ meson exchange, while for the charged $\rho$ meson
photoproduction we have shown that the $\rho$ meson exchange in the $t$-channel is
the dominant contribution. The dominance of the $t$-channel gives rise
to an angular distribution with a strong forward peak, in disagreement
with the presently available data.
For charged $\rho$ meson production a neutron target is necessary.
For this, we pointed out that application of the minimum momentum
spectator method may be used to achieve good kinematics in experiment.
The cross section was then shown to be dependent sensitively on the magnetic
moment of the $\rho$ meson and on the tensor coupling of
the $\rho$ meson with the nucleon
Such a study of $t$-channel dominant process is useful to obtain information of magnetic moment which
reflects the internal structure of an unstable particle.

\begin{acknowledgements}
The authors would also like to thank H.~Nagahiro for advice and helpful suggestions.
This work was supported partially by Industrialized Countries Instrument
 Education Cooperation Programme from Japan Student Service Organization.
A.H. is supported by the Grant-in-Aid for Scientific Research on
 Priority Areas ''Elucidation of New Hadrons with a Variety of Flavors''
 (E01: 21105006).
\end{acknowledgements}

\begin{thebibliography}{10}%
\makeatletter
\providecommand \@ifxundefined [1]{%
 \ifx #1\undefined \expandafter \@firstoftwo
 \else \expandafter \@secondoftwo
\fi
}%
\providecommand \@ifnum [1]{%
 \ifnum #1\expandafter \@firstoftwo
 \else \expandafter \@secondoftwo
\fi
}%
\providecommand \enquote [1]{``#1''}%
\providecommand \bibnamefont  [1]{#1}%
\providecommand \bibfnamefont [1]{#1}%
\providecommand \citenamefont [1]{#1}%
\providecommand\href[0]{\@sanitize\@href}%
\providecommand\@href[1]{\endgroup\@@startlink{#1}\endgroup\@@href}%
\providecommand\@@href[1]{#1\@@endlink}%
\providecommand \@sanitize [0]{\begingroup\catcode`\&12\catcode`\#12\relax}%
\@ifxundefined \pdfoutput {\@firstoftwo}{%
 \@ifnum{\z@=\pdfoutput}{\@firstoftwo}{\@secondoftwo}%
}{%
 \providecommand\@@startlink[1]{\leavevmode}%
 \providecommand\@@endlink[0]{}%
}{%
 \providecommand\@@startlink[1]{%
  \leavevmode
  \pdfstartlink
   attr{/Border[0 0 1 ]/H/I/C[0 1 1]}%
   user{/Subtype/Link/A<</Type/Action/S/URI/URI(#1)>>}%
  \relax
 }%
 \providecommand\@@endlink[0]{\pdfendlink}%
}%
\providecommand \url  [0]{\begingroup\@sanitize \@url }%
\providecommand \@url [1]{\endgroup\@href {#1}{\urlprefix}}%
\providecommand \urlprefix [0]{URL }%
\providecommand \Eprint[0]{\href }%
\@ifxundefined \urlstyle {%
  \providecommand \doi [1]{doi:\discretionary{}{}{}#1}%
}{%
  \providecommand \doi [0]{doi:\discretionary{}{}{}\begingroup
  \urlstyle{rm}\Url }%
}%
\providecommand \doibase [0]{http://dx.doi.org/}%
\providecommand \Doi[1]{\href{\doibase#1}}%
\providecommand \bibAnnote [3]{%
  \BibitemShut{#1}%
  \begin{quotation}\noindent
    \textsc{Key:}\ #2\\\textsc{Annotation:}\ #3%
  \end{quotation}%
}%
\providecommand \bibAnnoteFile [2]{%
  \IfFileExists{#2}{\bibAnnote {#1} {#2} {\input{#2}}}{}%
}%
\providecommand \typeout [0]{\immediate \write \m@ne }%
\providecommand \selectlanguage [0]{\@gobble}%
\providecommand \bibinfo [0]{\@secondoftwo}%
\providecommand \bibfield [0]{\@secondoftwo}%
\providecommand \translation [1]{[#1]}%
\providecommand \BibitemOpen[0]{}%
\providecommand \bibitemStop [0]{}%
\providecommand \bibitemNoStop [0]{.\EOS\space}%
\providecommand \EOS [0]{\spacefactor3000\relax}%
\providecommand \BibitemShut [1]{\csname bibitem#1\endcsname}%
\bibitem{Nambu:1997vw}%
  \BibitemOpen
  \bibfield{author}{%
  \bibinfo {author} {\bibfnamefont{Y.}~\bibnamefont{Nambu}},\ }%
  \bibfield{journal}{%
  \Doi{10.1103/PhysRev.106.1366}{\bibinfo {journal} {Phys. Rev.}}\ }%
  \textbf{\bibinfo {volume} {106}},\ \bibinfo {pages} {1366} (\bibinfo {year}
  {1957})%
  \bibAnnoteFile{NoStop}{Nambu:1997vw}%
\bibitem{Sakurai196906}%
  \BibitemOpen
  \bibfield{author}{%
  \bibinfo {author} {\bibfnamefont{J.~J.}\ \bibnamefont{Sakurai}},\ }%
  \emph{\bibinfo {title} {Currents and Mesons}}\ (\bibinfo {publisher} {Univ of
  Chicago Pr (Tx)},\ \bibinfo {year} {1969})\ ISBN \bibinfo {isbn}
  {9780226733838}%
  \bibAnnoteFile{NoStop}{Sakurai196906}%
\bibitem{Machleidt:1987hj}%
  \BibitemOpen
  \bibfield{author}{%
  \bibinfo {author} {\bibfnamefont{R.}~\bibnamefont{Machleidt}}, \bibinfo
  {author} {\bibfnamefont{K.}~\bibnamefont{Holinde}},\ and\ \bibinfo {author}
  {\bibfnamefont{C.}~\bibnamefont{Elster}},\ }%
  \bibfield{journal}{%
  \Doi{10.1016/S0370-1573(87)80002-9}{\bibinfo {journal} {Phys. Rept.}}\ }%
  \textbf{\bibinfo {volume} {149}},\ \bibinfo {pages} {1} (\bibinfo {year}
  {1987})%
  \bibAnnoteFile{NoStop}{Machleidt:1987hj}%
\bibitem{Bando:1984ej}%
  \BibitemOpen
  \bibfield{author}{%
  \bibinfo {author} {\bibfnamefont{M.}~\bibnamefont{Bando}}, \bibinfo {author}
  {\bibfnamefont{T.}~\bibnamefont{Kugo}}, \bibinfo {author}
  {\bibfnamefont{S.}~\bibnamefont{Uehara}}, \bibinfo {author}
  {\bibfnamefont{K.}~\bibnamefont{Yamawaki}},\ and\ \bibinfo {author}
  {\bibfnamefont{T.}~\bibnamefont{Yanagida}},\ }%
  \bibfield{journal}{%
  \Doi{10.1103/PhysRevLett.54.1215}{\bibinfo {journal} {Phys. Rev. Lett.}}\ }%
  \textbf{\bibinfo {volume} {54}},\ \bibinfo {pages} {1215} (\bibinfo {year}
  {1985})%
  \bibAnnoteFile{NoStop}{Bando:1984ej}%
\bibitem{Bando:1987br}%
  \BibitemOpen
  \bibfield{author}{%
  \bibinfo {author} {\bibfnamefont{M.}~\bibnamefont{Bando}}, \bibinfo {author}
  {\bibfnamefont{T.}~\bibnamefont{Kugo}},\ and\ \bibinfo {author}
  {\bibfnamefont{K.}~\bibnamefont{Yamawaki}},\ }%
  \bibfield{journal}{%
  \Doi{10.1016/0370-1573(88)90019-1}{\bibinfo {journal} {Phys. Rept.}}\ }%
  \textbf{\bibinfo {volume} {164}},\ \bibinfo {pages} {217} (\bibinfo {year}
  {1988})%
  \bibAnnoteFile{NoStop}{Bando:1987br}%
\bibitem{Sakai:2004cn}%
  \BibitemOpen
  \bibfield{author}{%
  \bibinfo {author} {\bibfnamefont{T.}~\bibnamefont{Sakai}}\ and\ \bibinfo
  {author} {\bibfnamefont{S.}~\bibnamefont{Sugimoto}},\ }%
  \bibfield{journal}{%
  \Doi{10.1143/PTP.113.843}{\bibinfo {journal} {Prog. Theor. Phys.}}\ }%
  \textbf{\bibinfo {volume} {113}},\ \bibinfo {pages} {843} (\bibinfo {year}
  {2005}),\ \Eprint{http://arxiv.org/abs/hep-th/0412141}{arXiv:hep-th/0412141}%
  \bibAnnoteFile{NoStop}{Sakai:2004cn}%
\bibitem{Sakai:2005yt}%
  \BibitemOpen
  \bibfield{author}{%
  \bibinfo {author} {\bibfnamefont{T.}~\bibnamefont{Sakai}}\ and\ \bibinfo
  {author} {\bibfnamefont{S.}~\bibnamefont{Sugimoto}},\ }%
  \bibfield{journal}{%
  \Doi{10.1143/PTP.114.1083}{\bibinfo {journal} {Prog. Theor. Phys.}}\ }%
  \textbf{\bibinfo {volume} {114}},\ \bibinfo {pages} {1083} (\bibinfo {year}
  {2005}),\ \Eprint{http://arxiv.org/abs/hep-th/0507073}{arXiv:hep-th/0507073}%
  \bibAnnoteFile{NoStop}{Sakai:2005yt}%
\bibitem{Roca:2005nm}%
  \BibitemOpen
  \bibfield{author}{%
  \bibinfo {author} {\bibfnamefont{L.}~\bibnamefont{Roca}}, \bibinfo {author}
  {\bibfnamefont{E.}~\bibnamefont{Oset}},\ and\ \bibinfo {author}
  {\bibfnamefont{J.}~\bibnamefont{Singh}},\ }%
  \bibfield{journal}{%
  \Doi{10.1103/PhysRevD.72.014002}{\bibinfo {journal} {Phys. Rev.}}\ }%
  \textbf{\bibinfo {volume} {D72}},\ \bibinfo {pages} {014002} (\bibinfo {year}
  {2005}),\ \Eprint{http://arxiv.org/abs/hep-ph/0503273}{arXiv:hep-ph/0503273}%
  \bibAnnoteFile{NoStop}{Roca:2005nm}%
\bibitem{Khemchandani:2011et}%
  \BibitemOpen
  \bibfield{author}{%
  \bibinfo {author} {\bibfnamefont{K.}~\bibnamefont{Khemchandani}}, \bibinfo
  {author} {\bibfnamefont{H.}~\bibnamefont{Kaneko}}, \bibinfo {author}
  {\bibfnamefont{H.}~\bibnamefont{Nagahiro}},\ and\ \bibinfo {author}
  {\bibfnamefont{A.}~\bibnamefont{Hosaka}},\ }%
  \bibfield{journal}{%
  \Doi{10.1103/PhysRevD.83.114041}{\bibinfo {journal} {Phys.Rev.}}\ }%
  \textbf{\bibinfo {volume} {D83}},\ \bibinfo {pages} {114041} (\bibinfo {year}
  {2011}),\ \Eprint{http://arxiv.org/abs/1104.0307}{arXiv:1104.0307 [hep-ph]}%
  \bibAnnoteFile{NoStop}{Khemchandani:2011et}%
\bibitem{Khemchandani:2011mf}%
  \BibitemOpen
  \bibfield{author}{%
  \bibinfo {author} {\bibfnamefont{K.}~\bibnamefont{Khemchandani}}, \bibinfo
  {author} {\bibfnamefont{A.}~\bibnamefont{Torres}}, \bibinfo {author}
  {\bibfnamefont{H.}~\bibnamefont{Kaneko}}, \bibinfo {author}
  {\bibfnamefont{H.}~\bibnamefont{Nagahiro}},\ and\ \bibinfo {author}
  {\bibfnamefont{A.}~\bibnamefont{Hosaka}},\ }%
  \bibfield{journal}{%
  \Doi{10.1103/PhysRevD.84.094018}{\bibinfo {journal} {Phys. Rev. D}}\ }%
  \textbf{\bibinfo {volume} {84}},\ \bibinfo {pages} {094018} (\bibinfo {month}
  {Nov}\ \bibinfo {year} {2011}),\
  \url{http://link.aps.org/doi/10.1103/PhysRevD.84.094018}%
  \bibAnnoteFile{NoStop}{Khemchandani:2011mf}%
\bibitem{Wu:2005wf}%
  \BibitemOpen
  \bibfield{author}{%
  \bibinfo {author} {\bibfnamefont{C.}~\bibnamefont{Wu}} \emph{et~al.},\ }%
  \bibfield{journal}{%
  \Doi{10.1140/epja/i2004-10093-9}{\bibinfo {journal} {Eur. Phys. J.}}\ }%
  \textbf{\bibinfo {volume} {A23}},\ \bibinfo {pages} {317} (\bibinfo {year}
  {2005})%
  \bibAnnoteFile{NoStop}{Wu:2005wf}%
\bibitem{:1968ke}%
  \BibitemOpen
  \bibfield{author}{%
  \bibinfo {author} {\bibfnamefont{A.-B.-B.-H.-H.-M.}\
  \bibnamefont{Collaboration}} (\bibinfo {collaboration}
  {Aachen-Berlin-Bonn-Hamburg-Hedielberg-Munich}),\ }%
  \bibfield{journal}{%
  \Doi{10.1103/PhysRev.175.1669}{\bibinfo {journal} {Phys. Rev.}}\ }%
  \textbf{\bibinfo {volume} {175}},\ \bibinfo {pages} {1669} (\bibinfo {year}
  {1968})%
  \bibAnnoteFile{NoStop}{:1968ke}%
\bibitem{Friman:1995qm}%
  \BibitemOpen
  \bibfield{author}{%
  \bibinfo {author} {\bibfnamefont{B.}~\bibnamefont{Friman}}\ and\ \bibinfo
  {author} {\bibfnamefont{M.}~\bibnamefont{Soyeur}},\ }%
  \bibfield{journal}{%
  \Doi{10.1016/0375-9474(96)00011-5}{\bibinfo {journal} {Nucl. Phys.}}\ }%
  \textbf{\bibinfo {volume} {A600}},\ \bibinfo {pages} {477} (\bibinfo {year}
  {1996}),\
  \Eprint{http://arxiv.org/abs/nucl-th/9601028}{arXiv:nucl-th/9601028}%
  \bibAnnoteFile{NoStop}{Friman:1995qm}%
\bibitem{Fujiwara:1984mp}%
  \BibitemOpen
  \bibfield{author}{%
  \bibinfo {author} {\bibfnamefont{T.}~\bibnamefont{Fujiwara}}, \bibinfo
  {author} {\bibfnamefont{T.}~\bibnamefont{Kugo}}, \bibinfo {author}
  {\bibfnamefont{H.}~\bibnamefont{Terao}}, \bibinfo {author}
  {\bibfnamefont{S.}~\bibnamefont{Uehara}},\ and\ \bibinfo {author}
  {\bibfnamefont{K.}~\bibnamefont{Yamawaki}},\ }%
  \bibfield{journal}{%
  \Doi{10.1143/PTP.73.926}{\bibinfo {journal} {Prog. Theor. Phys.}}\ }%
  \textbf{\bibinfo {volume} {73}},\ \bibinfo {pages} {926} (\bibinfo {year}
  {1985})%
  \bibAnnoteFile{NoStop}{Fujiwara:1984mp}%
\bibitem{Davidson:2001rk}%
  \BibitemOpen
  \bibfield{author}{%
  \bibinfo {author} {\bibfnamefont{R.~M.}\ \bibnamefont{Davidson}}\ and\
  \bibinfo {author} {\bibfnamefont{R.}~\bibnamefont{Workman}},\ }%
  \bibfield{journal}{%
  \Doi{10.1103/PhysRevC.63.025210}{\bibinfo {journal} {Phys. Rev.}}\ }%
  \textbf{\bibinfo {volume} {C63}},\ \bibinfo {pages} {025210} (\bibinfo {year}
  {2001}),\
  \Eprint{http://arxiv.org/abs/nucl-th/0101066}{arXiv:nucl-th/0101066}%
  \bibAnnoteFile{NoStop}{Davidson:2001rk}%
\bibitem{ThomasWeise200105}%
  \BibitemOpen
  \bibfield{author}{%
  \bibinfo {author} {\bibfnamefont{A.~W.}\ \bibnamefont{Thomas}}\ and\ \bibinfo
  {author} {\bibfnamefont{W.}~\bibnamefont{Weise}},\ }%
  \emph{\bibinfo {title} {The Structure of the Nucleon}},\ \bibinfo {edition}
  {1st}\ ed.\ (\bibinfo {publisher} {Wiley-VCH},\ \bibinfo {year} {2001})\ ISBN
  \bibinfo {isbn} {9783527402977}%
  \bibAnnoteFile{NoStop}{ThomasWeise200105}%
\bibitem{Soding:1965nh}%
  \BibitemOpen
  \bibfield{author}{%
  \bibinfo {author} {\bibfnamefont{P.}~\bibnamefont{Soding}},\ }%
  \bibfield{journal}{%
  \Doi{10.1016/0031-9163(66)90451-3}{\bibinfo {journal} {Phys. Lett.}}\ }%
  \textbf{\bibinfo {volume} {19}},\ \bibinfo {pages} {702} (\bibinfo {year}
  {1966})%
  \bibAnnoteFile{NoStop}{Soding:1965nh}%
\bibitem{Ross:1965qa}%
  \BibitemOpen
  \bibfield{author}{%
  \bibinfo {author} {\bibfnamefont{M.~H.}\ \bibnamefont{Ross}}\ and\ \bibinfo
  {author} {\bibfnamefont{L.}~\bibnamefont{Stodolsky}},\ }%
  \bibfield{journal}{%
  \Doi{10.1103/PhysRev.149.1172}{\bibinfo {journal} {Phys.Rev.}}\ }%
  \textbf{\bibinfo {volume} {149}},\ \bibinfo {pages} {1172} (\bibinfo {year}
  {1966})%
  \bibAnnoteFile{NoStop}{Ross:1965qa}%
\bibitem{Clark:1970xr}%
  \BibitemOpen
  \bibfield{author}{%
  \bibinfo {author} {\bibfnamefont{R.~B.}\ \bibnamefont{Clark}},\ }%
  \bibfield{journal}{%
  \Doi{10.1103/PhysRevD.1.2152}{\bibinfo {journal} {Phys. Rev.}}\ }%
  \textbf{\bibinfo {volume} {D1}},\ \bibinfo {pages} {2152} (\bibinfo {year}
  {1970})%
  \bibAnnoteFile{NoStop}{Clark:1970xr}%
\bibitem{Glozman:2011gf}%
  \BibitemOpen
  \bibfield{author}{%
  \bibinfo {author} {\bibfnamefont{L.}~\bibnamefont{Glozman}}, \bibinfo
  {author} {\bibfnamefont{C.}~\bibnamefont{Lang}},\ and\ \bibinfo {author}
  {\bibfnamefont{M.}~\bibnamefont{Limmer}},\ }%
  \bibfield{journal}{%
  \Doi{10.1016/j.physletb.2011.09.102}{\bibinfo {journal} {Phys.Lett.}}\ }%
  \textbf{\bibinfo {volume} {B705}},\ \bibinfo {pages} {129} (\bibinfo {year}
  {2011}),\ \Eprint{http://arxiv.org/abs/1106.1010}{arXiv:1106.1010 [hep-ph]}%
  \bibAnnoteFile{NoStop}{Glozman:2011gf}%
\bibitem{Hilpert:1970jd}%
  \BibitemOpen
  \bibfield{author}{%
  \bibinfo {author} {\bibfnamefont{H.~G.}\ \bibnamefont{Hilpert}}
  \emph{et~al.},\ }%
  \bibfield{journal}{%
  \Doi{10.1016/0550-3213(70)90464-5}{\bibinfo {journal} {Nucl. Phys.}}\ }%
  \textbf{\bibinfo {volume} {B21}},\ \bibinfo {pages} {93} (\bibinfo {year}
  {1970})%
  \bibAnnoteFile{NoStop}{Hilpert:1970jd}%
\bibitem{Nakano:2008ee}%
  \BibitemOpen
  \bibfield{author}{%
  \bibinfo {author} {\bibfnamefont{T.}~\bibnamefont{Nakano}} \emph{et~al.}
  (\bibinfo {collaboration} {LEPS Collaboration}),\ }%
  \bibfield{journal}{%
  \Doi{10.1103/PhysRevC.79.025210}{\bibinfo {journal} {Phys.Rev.}}\ }%
  \textbf{\bibinfo {volume} {C79}},\ \bibinfo {pages} {025210} (\bibinfo {year}
  {2009}),\ \Eprint{http://arxiv.org/abs/0812.1035}{arXiv:0812.1035 [nucl-ex]}%
  \bibAnnoteFile{NoStop}{Nakano:2008ee}%
\end{thebibliography}
%

\end{document}